\documentclass[pdflatex,sn-chicago]{sn-jnl}

\jyear{2021}%

\theoremstyle{thmstyleone}%
\theoremstyle{thmstyletwo}%

\theoremstyle{thmstylethree}%

\begin{document}

\title[Article Title]{Multi-dimensional Racism Classification during COVID-19: Stigmatization, Offensiveness, Blame, and Exclusion}


\author*[1]{\fnm{Xin} \sur{ Pei}}\email{xin.pei@unimelb.edu.au}
\equalcont{These authors contributed equally to this work.}

\author[2]{\fnm{Deval} \sur{Mehta}}\email{Deval.mehta1092@gmail.com}
\equalcont{These authors contributed equally to this work.}

\affil*[1]{\orgdiv{School of Culture and Communication}, \orgname{The University of Melbourne}, \orgaddress{\street{Parkville}, \city{Melbourne}, \postcode{3010}, \state{VIC}, \country{Australia}}}

\affil[2]{\orgdiv{Monash eResearch Center}, \orgname{Monash University}, \orgaddress{\street{Clayton}, \city{Melbourne}, \postcode{3800}, \state{VIC}, \country{Australia}}}


\abstract{Transcending the binary categorization of racist texts, our study takes cues from social science theories to develop a multi-dimensional model for racism detection, namely stigmatization, offensiveness, blame, and exclusion. With the aid of BERT and topic modeling, this categorical detection enables insights into the underlying subtlety of racist discussion on digital platforms during COVID-19. Our study contributes to enriching the scholarly discussion on deviant racist behaviours on social media. First, a stage-wise analysis is applied to capture the dynamics of the topic changes across the early stages of COVID-19 which transformed from a domestic epidemic to an international public health emergency and later to a global pandemic. Furthermore, mapping this trend enables a more accurate prediction of public opinion evolvement concerning racism in the offline world, and meanwhile, the enactment of specified intervention strategies to combat the upsurge of racism during the global public health crisis like COVID-19. In addition, this interdisciplinary research also points out a direction for future studies on social network analysis and mining. Integration of social science perspectives into the development of computational methods provides insights into more accurate data detection and analytics.}




\maketitle

\section{Introduction}\label{sec1}

The global outbreak of COVID-19 has been accompanied by the worldwide upsurge of racism. An increasing research stream has illuminated the more infectious nature of racist reactions than coronavirus, which is leading towards a more harmful social consequence \citep{kapilashrami2020mental,wang2021m}. BBC has reported that the United Nations raised racially motivated violence and other hate incidents against Asian Americans to "an alarming level" in 2020\citep{BBCR}. And hate crimes occurred in the New York City in 2020 experienced a nine-fold increase from the previous year.Therefore, it has become urgent to comprehend the racist discourse so as to enact effective intervention strategies to prevent the escalation of deviant behaviours such as hate crimes and social exclusion during COVID-19. 

Against this backdrop, many studies have drawn attention to social media platforms which provide critical avenues for pandemic-related public discussion. Scholars have widely adopted highly advanced computational methods and state-of-the-art language models for big social data analytics on these platforms, with the purpose of achieving a better understanding of racist reactions from the public. Unsupervised machine learning techniques such as topic modeling, keyword clustering have been widely employed in studies (e.g., \citep{tahmasbi2021go}) for analysing Twitter and Reddit data during COVID-19. Scholars \citep{he2021racism,lu2020fear} have also adopted supervised learning methods such as Support Vector Machines(SVMs) and Transformers for hate and racist speech detection.

Despite the contribution in technical advancement, the extant literature shows the tendency of neglecting the theoretical foundation for data detection and analysis - that is how to define racism in the first place. To specify, the existing computational techniques and models tend to apply a binary definition that primarily categorises the linguistic features of texts into either the racist or non-racist ones (e.g.,~\citep{he2021racism}). It is important to note that some studies mentioned different dimensions for racism identification. For instance, study by Davidson and colleagues (2017) utilized offensive language for automatic detection of hate speech. In the same vein, some other studies \citep{fan2020stigmatization,liu2022deep} particularly focused on stigmatization. However, they are not only restricted in numbers but also lack of a comprehensive model based on a summary of relevant indicators from social science studies. Given the dynamic nature of racist behaviours \citep{richeson2018psychology}, a comprehensive classification capturing more nuances of racist discourse will allow for more insights into the behavioural change across different stages of COVID-19. 

To fill this research gap, our study transcends the binary of (non)racism by introducing a model that classifies racist behaviours into four categories - stigmatization, offensiveness, blame, and exclusion. It is important to note that this model is built upon a combination of social science theories and computational methods. To specify, while the categorization is generated from prior scholarly discussion on racism across the domains of sociology, psychology, and social psychology, the application of the model involves deep learning techniques - BERT (Bi-directional Encoder Representations from Transformers) \citep{devlin2018bert} and topic modelling \citep{blei2003latent}.

Our study makes unique contribution that enriches the scholarly discussion on deviant racist behaviours on social media. First, applying this model on a stage-wise analysis, our study captures the dynamic evolvement of racist behaviours across the early development of COVID-19 – how the four racist categories competed with one another at different stages, and how the themes of each category shifted across time. Furthermore, mapping this trend will enable a more accurate prediction of public opinion evolvement concerning racism in the offline world, and meanwhile, the enactment of specified intervention strategies to combat the upsurge of racism during the global public health crisis like COVID-19. In addition, this interdisciplinary research also enhances a direction for future studies on social network analysis and mining. Integration of social science perspectives into the development of computational methods can provide a new route for a more accurate data detection and analytics.

\section{Literature review}\label{sec2}

\subsection{Racism, social media, and COVID-19}\label{subsec21}
Many have argued that social media platforms are providing critical avenues for racist opinion expression. Especially, the widely advocated speech freedom on social media platforms is paving the way for toxic and provocative languages replete with trolls, often with the target at a particular race/ethnicity, nation, or (im)migrant community \citep{lim2017freedom}. Anonymity further enables hate speech and biased opinions to avoid detection \citep{keum2018racism}. Meanwhile, boundless connectivity on social media platforms allows racist opinions to travel at a fast speed and to reach a broad scope of audiences \citep{he2021racism}. Moreover, such connectivity also permits people with similar racial ideologies to cluster and collectively build up the racist discourse to increase its visibility and influence online \citep{kapilashrami2020mental}. Therefore, \cite{matamoros2017platformed} coined the term “platformed” racism to refer to people’s usage of affordances on different social media platforms to duplicate and extend the offline social inequalities. \citep{oboler2016measuring} indicated the emergence of ‘Hate 2.0”, under which the repetitive occurrence of hate speech on social media keeps on justifying the racist discourse as a normalized collective behaviour.

It is important to note the rise of racism on social media during COVID-19. As one of the most severe global pandemics since the turn of the new millennium, COVID-19 has caused more than forty million confirmed cases and almost five million deaths across the globe till the submission date of this manuscript. Suspected to be originated from Wuhan, China, the global outbreak of COVID-19 has widely raised social exclusion against China, which has been evolved into discrimination, bias, and even hatred against Chinese and even Asians at large. This phenomenon has been unofficially coined as sinophobia. 

It is worth noting that, racism has been largely extended to the online world under the pandemic. On 16th March 2020, a post from the official Twitter account of Donald Trump, the former president of the United States of America, referred to COVID-19 as Chinese virus. Ironically, this overtly racist and xenophobic label immediately became an emergent popular hashtag - \texttt{\#}chinesevirus, which was massively disseminated and circulated on Twitter and other social media platforms. Besides \texttt{\#}chinesevirus, social media platforms have witnessed the proliferation of many other offensive hashtags centring on a particular race and nation in the pandemic context, such as \texttt{\#}kungflu embodying the conflation of coronavirus with racial/ethnic cultural identities, and \texttt{\#}boycottchina manifesting social exclusion. Hate speech and discriminative opinions are massively circulated and disseminated through these hashtags. Consequently, many scholars have initiated the investigation into racist reactions on social media during COVID-19. The following section will elaborate on the contribution and drawbacks of the research stream using computational methods.

\subsection{Bridging social science theory and computational methods}\label{subsec22}
Many studies have adopted computational methods for big data mining and social network analytics to better understand the dynamics of the racist deviant behaviors on social media platform from a macro-level. For example, \citep{garland2020countering} used the supervised machine learning models such as Random Forests and Support Vector Machines (SVMs) for classification of hate speech. Leveraging the large textual data corpus, studies have also used unsupervised techniques such as word2vec~\citep{mikolov2013efficient}, Glove embeddings~\citep{pennington2014glove} for topic clustering and keywords analysis. More recently, with the advent of deep learning and availability of training data, models such as Long-Short Term Memory Networks (LSTMs)~\citep{hochreiter1997long}, RNNs, and much recently Transformers such as BERT~\citep{devlin2018bert} have been employed in most studies~\citep{he2021racism,lu2020fear} for classification of textual data on social media platforms.

Regardless of the contribution made to advancing the tools and techniques, prior studies tend to ignore the most fundamental issue that shall be addressed in the first place – that is how to define racism. Especially, many studies tend to adopt a binary classification of linguistic features that categorizes the texts into either racist or non-racist ones. Although some studies have committed the efforts to enriching the linguistic features of hateful and offensive speech \citep{abderrouaf2019online,fahim2021detecting}, the extant classification still tends to largely underestimate the complexity of racist behaviors, thereby leading to an oversimplified mechanism for racist data detection and analysis. This tends to prevent the discovery of the nuances of the themes embodied in the racist opinion expression, and the dynamics of the themes that are very likely to evolve alongside the development of a public event \citep{pei2020coronavirus}. 

To fill this research gap, our study proposes a multi-dimensional model to detect and classify racism which is built upon the conceptualization of racism in prior social science research. To specify, transcending the binary category, our model specifies racist behaviors into stigmatization, offensiveness, blame, and exclusion. Taking references from the study by \citep{miller2001theoretical} centering on theorizing stigma, our model defines stigma as confirming negative stereotypes for conveying a devalued social identity within a particular context. Similarly, built upon the research by \citep{jeshion2013expressivism} surrounding the expression of offensive slurs, we refer to offensiveness as attacking a particular social group through aggressive and abusive language. The study by \citep{coombs2000empirical} in the context of Texaco’s racism crisis points a direction for framing blame as attributing the responsibility for the negative consequences of the crisis to one social group. The dimension of exclusion stems from the study by \citep{bailey2005racialised} that noted exclusion as a critical step of racializing others which embodies the process of othering to draw a clear boundary between in-group and out-group members. Please refer Table \ref{tab1} which includes the definition of the four dimensions accompanied by the corresponding examples from the dataset. 

This multi-dimensional classification model is applied to a stage-wise analysis, with the purpose of mapping the dynamics – how these four racist themes were competing with one another alongside the development of COVID-19. This will provide a more nuanced idea about the trend regarding the possibly shifting focus of public opinion concerning racism. Especially, we focus on the most turbulent early phase of COVID-19 (Jan to Apr 2020) where the unexpected and constant global expansion of the virus kept on changing people’s perception of this public health crisis and how it is related to race and nationality. To specify, this research divides the early phase into three stages based on the changing definitions of COVID-19 made by the World Health Organization (WHO) - (1) 1st to 31st Jan 2020 as a domestic epidemic referred to as stage 1 (S1); (2) 1st Feb to 11th Mar 2020 as an International Public Health Emergency (after the announcement made by WHO on 1st Feb) referred to as stage 2 (S2); (3) 12th Mar to 30th Apr 2020 as a global pandemic (based on the new definition given by WHO on 11th Mar) referred to as stage 3 (S3). We select Twitter, the most influential platform for political online discussion, as the field for data mining and analysis.

\section{Data and Methods}\label{sec3}

This section deals with five parts - first, it outlines method used to scrape the data; second, it defines the four dimensions of racism; third, it describes the process of annotation; fourth, it explains the method employed for category-based racism and xenophobia detection; and last, it details the process of topic modelling employed for extracting topics from the categorized data.

Dataset of this research is comprised of 247,153 tweets extracted through Tweepy API\footnote{https://www.tweepy.org/}. We built a custom python-based wrapper utilizing the Tweepy API functionalities to continuously scrape the data starting from the 1st January until the 30th April 2020, which falls within our interest period of early covid-19 including the three durations of a domestic epidemic, an International Public Health Emergency, and eventually a global pandemic as highlighted in the paper.

For selecting the hashtags to scrape the Twitter data we first used the most common and topmost hashtags - \texttt{\#}chinesevirus and \texttt{\#}chinavirus, which were also used by the other popular studies \citep{tahmasbi2021go,ziems2020racism} of analysing covid-19 data on Twitter. For selecting all the other hashtags in our data scraping process, we followed a dynamic hashtag selection process which is a similar approach to the study \citep{srikanthfinding} relevant for “rapidly-evolving online datasets”.  Our strategy of dynamic hashtag scraping involves the following steps:

\begin{itemize}
    \item After scraping a sample of ~500 tweets from the topmost hashtags, we collect the most frequent top 5 hashtags occurring in those samples of 500 tweets.
    \item 	The collected top 5 most frequent hashtags (excluding the ones which were used to collect them) are then used to scrape new tweet samples containing them.
    \item We then repeat the first step for the new sample of the scraped tweets and collect the new most frequent top 5 hashtags having at least ~50 occurrences.
    \item 	In the above step, we noticed that most of the new hashtags don’t have a high frequency of repetition, so we stop the recursive scraping involving the first step at this stage.
\end{itemize}

The above process is adopted for the first two weeks at the beginning of a new stage of the data collection in our duration of scraping i.e., this strategy is first employed on the 1st January 2020 (beginning of stage 1); then repeated on the 1st February 2020 (beginning of stage 2) and on 12th March 2020 (beginning of stage 3).

\begin{figure}[h]%
\centering
\includegraphics[width=0.9\textwidth]{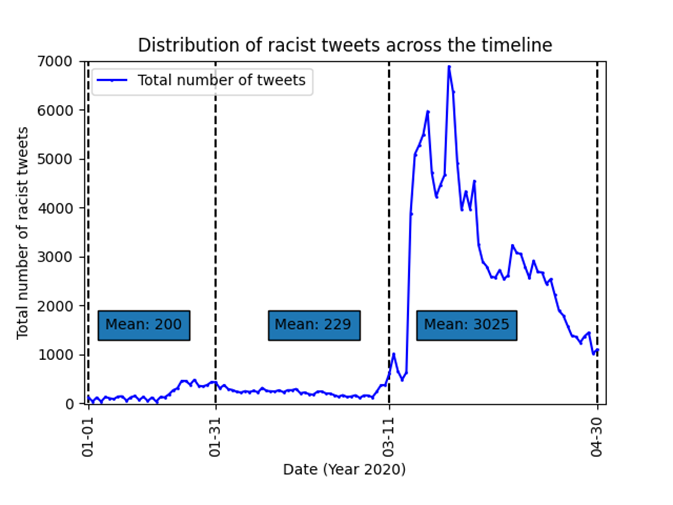}
\caption{Analysis of the number of tweets returned daily by our custom wrapper (based on Tweepy API) from 1st Jan to 30th Apr, 2020.}\label{fig1}
\end{figure}

With this strategy, we developed the following list of hashtags which were then used to mine the data - \texttt{\#}chinavirus, \texttt{\#}chinesevirus, \texttt{\#}boycottchina, \texttt{\#}ccpvirus, \texttt{\#}chinaflu, \texttt{\#}china\_is\_terrorist, \texttt{\#}chinaliedandpeopledied, \texttt{\#}chinaliedpeopledied, \texttt{\#}chinalies, \texttt{\#}chinamustpay, \texttt{\#}chinapneumonia, \texttt{\#}chinazi, \texttt{\#}chinesebioterrorism, \texttt{\#}chinesepneumonia, \texttt{\#}chinesevirus19, \texttt{\#}chinesewuhanvirus, \texttt{\#}viruschina, and  \texttt{\#}wuflu. The extracted tweets from the above hashtags are further divided into three stages that define the early development of Covid-19 as mentioned earlier. We show the number of tweets extracted for each day using this method in Fig~\ref{fig1}.

\subsection{Method}
\subsubsection{Category-based racism and xenophobia detection}
Beyond a binary categorization of racism and xenophobia, this research applies the perspective of social science to categorizing racism and xenophobia into four dimensions as demonstrated in Table~\ref{tab1}. This basically translates into a problem of five class classification of text data, where four classes represent the four types of racism, and the fifth class refers to the category of non-racist and nonxenophobic.

\begin{table}[h]
\caption{Definition and example of categorization of racist and xenophobic behaviors.}\label{tab1}%
\resizebox{\textwidth}{!}{\begin{tabular}{l|p{80mm}p{80mm}}
\hline
Category & Definition & Example \\
\hline
Stigmatization & Confirming negative stereotypes for conveying \newline a devalued social identity within a particular context\citep{miller2001theoretical} & “For all the \#ChinaVirus jumped from a bat at the wet market”\\
\hline
Offensiveness & Attacking a particular social group \newline through aggressive and abusive language \newline \citep{jeshion2013expressivism} & “Real misogyny in communist China. \#chinazi \#China\_is\_terrorist \#China\_is\_terrorists \#FuckTheCCP” \\
\hline
Blame & Attributing the responsibility for the \newline negative consequences of the crisis to one social group \newline \citep{coombs2000empirical} & “These Chinese are absolutely disgusting. They spread the \#ChineseVirus. Their lies created a pandemic \#ChinaMustPay” \\
\hline
Exclusion & the process of othering to draw a clear boundary \newline between in-group and out-group members \newline \citep{bailey2005racialised} & “China deserves to be isolated by all means forever. SARS was also initiated in China, 2003 by eating anything \& everything \#BoycottChina” \\
\hline
\end{tabular}}
\end{table}

\subsubsection{Annotated dataset}
To train machine learning and deep learning classifiers for this task, we aimed to build a reasonable size dataset (not so large that it becomes difficult to annotate, and not so small enough to compromise on proper training and evaluation of our methods). Thus, we selected 6,000 as the number of representative tweets that can be utilized for training the model. To evenly represent the opinions from the three stages, we divided the selection of 6000 tweets into 2000 tweets from each stage (S1, S2, S3). These 2000 tweets were then randomly selected from each development stage.

We hired four research assistants who were initially trained under the supervision of one co-author on a pilot data of 200 tweets to categorize them into four different categories based on their definitions. The annotation followed a coding method with 0 representing stigmatization, 1 for offensiveness, 2 for blame, and 3 for exclusion in alignment with the linguistic features of the tweets. The non-marked tweets were regarded as non-racist and non-xenophobic and represented class category 4. We limited the annotation for each tweet to only one label which aligned with the strongest category. All four research assistants had Asian ethnicity (Chinese).

After completing their initial training and review from the co-author, they were provided feedback if there was a dispute in labeling. The four research assistants reached overall inter-coder reliability above 70\%, which is a moderate to high threshold selected by prior studies \citep{guntuku2019understanding,jaidka2019brevity} for reliability of data annotation. Post this pilot data training of the four research assistants, they were then given 500 tweets from each stage to categorize, which enabled us to make our dataset of 6000 tweets across the three stages. The distribution of 6000 tweets amongst the five classes is as follows - 1318 stigmatization, 1172 offensive, 1045 blame, 1136 exclusion, and 1329 non-racist and non-xenophobic.

We view the task of classification of the above-mentioned categories as a supervised learning problem and target developing machine learning and deep learning techniques for the same. We firstly pre-process the input data text by removing punctuation and URLs from a text sample and converting it to lower case before providing it to train our models. We split the data into random train and test splits with 90:10 ratio for training and evaluating the performance of our models respectively by using the standard five-fold cross validation.

\subsection{BERT}
\begin{figure}[h]%
\centering
\includegraphics[width=0.9\textwidth]{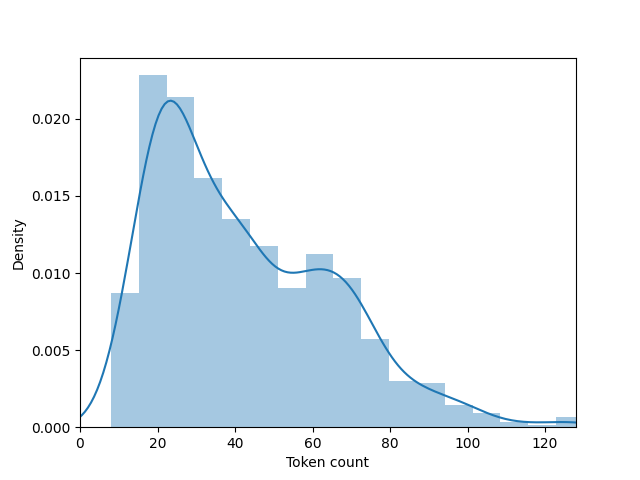}
\caption{Density distribution of token lengths of the tweets in our dataset.}\label{fig2}
\end{figure}

Recently, word language models such as Bi-directional Encoder Representations from Transformers (BERT) \citep{devlin2018bert} have become extremely popular due to their state-of-the-art performance on natural language processing tasks. Due to the nature of bi-directional training of BERT, it can learn the word representations from unlabelled text data powerfully and enables it to have a better performance compared to the other machine learning and deep learning techniques \citep{devlin2018bert}. The common approach for adopting BERT for a specific task on a smaller dataset is to fine-tune a pre-trained BERT model which has already learnt the deep context-dependent representations. We select the “bert-base-uncased” model which comprises of 12 layers, 12 self-attention heads, a hidden size of 768 totalling 110M parameters. We fine-tune the BERT model with a categorical cross-entropy loss for the five categories. The various hyperparameters used for fine-tuning the BERT model are selected as recommended from the paper \citep{devlin2018bert}. We use the AdamW optimizer with the standard learning rate of 2e-5, a batch size of 16, and train it for 5 epochs. For selecting the maximum length of the sequences, we tokenize the whole dataset using Bert tokenizer and check the distribution of the token lengths. We notice that the minimum value of token length is 8, maximum is 130, median is 37 and mean is ~42. Based on the density distribution shown in Figure~\ref{fig2}, we experiment with two values of sequence length – 64 and 128 and find that the sequence length of 64 provides a better performance.

As additional baselines, we also train two more techniques. Long Short Term Memory Networks (LSTMs) \citep{hochreiter1997long} have been very popular with text data as they can learn the dependencies of various words in the context of a text. Also, machine learning algorithms such as Support Vector Machine (SVMs) \citep{hearst1998support} have been used previously by researchers for text classification tasks. Moreover, the use of various feature extraction techniques such as term frequency inverse document frequency (TF-IDF), word2vec and Bag-of-Words (BoW) has proven to improve the performance of the classifiers \citep{li2019key,zhang2010understanding}. The work in \citep{gebre2013improving} explored the use of TF-IDF on machine learning classifiers such as SVM and it was found that it helped to improve the classification performance significantly compared to the original baselines. It was found that TF-IDF and BoW perform the best with uni-gram or bi-gram collection of word features. We operate the bi-gram BoW, bi-gram TF-IDF and Word2Vec feature engineering techniques for training our SVM model. We adopt the same data pre-processing and implementation technique as mentioned earlier and train the SVM with grid search, a 5-layer LSTM (using the pre-trained Glove \citep{pennington2014glove} embeddings) and BERT model for the category detection of the racist and xenophobic tweets.

\begin{table}[!t]
\centering
\caption{Performance of different models on the manually annotated test dataset. Mean accuracy and f1-score for the five-folds}\label{tab2}%
\resizebox{0.65\textwidth}{!}{\begin{tabular}{l|cc}
\hline
Technique & Accuracy(\%) & F1-score \\
\hline
SVM & 69.04 & 0.66 \\
SVM + TF-IDF Features & 75.8 & 0.74 \\
SVM + BOW Features & 71.6 & 0.70 \\
SVM + Word2Vec Features & 71.4 & 0.70 \\
LSTM & 74.01 & 0.72 \\
BERT & \textbf{86.06} & \textbf{0.81} \\
\hline
\end{tabular}}
\end{table}

\begin{figure}[h]%
\centering
\includegraphics[width=0.9\textwidth]{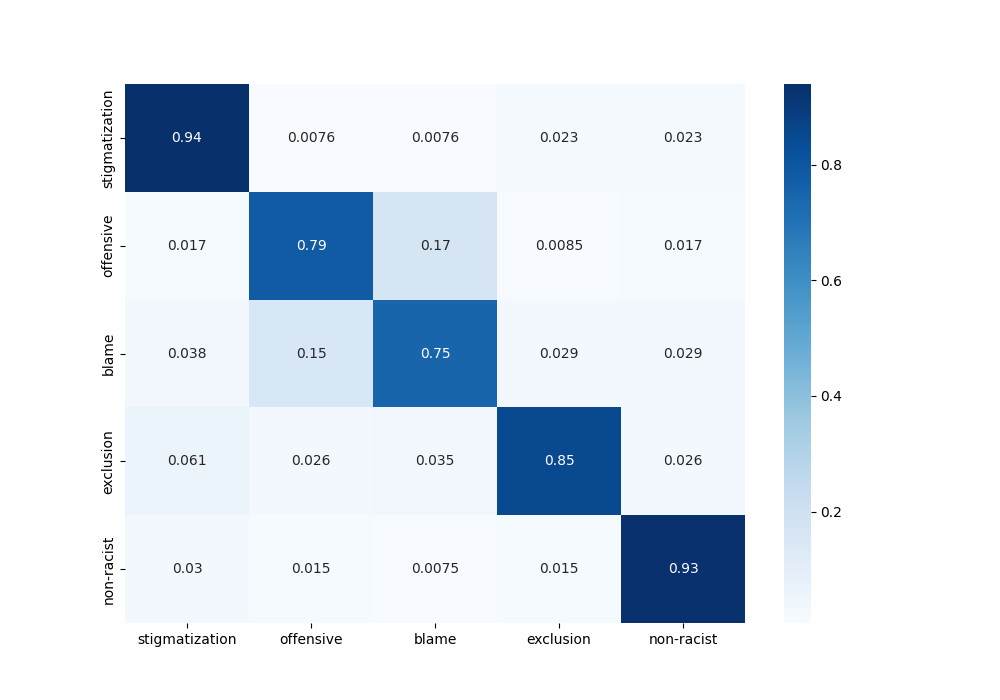}
\caption{Confusion matrix (averaged for the 5-folds of validation data) of our trained BERT model for racism classification.}\label{fig3}
\end{figure}

For evaluating the machine learning and deep learning approaches on our test dataset, we use the metrics of average accuracy and weighted f1-score for the five categories. The performance of the models is shown in Table~\ref{tab2}. It can be seen from Table~\ref{tab2} that the fine-tuned BERT model performs the best compared to SVM and LSTM in terms of both accuracy and f1 score. Although adding engineered features from TF-IDF improves the performance of the SVM classifier significantly, it cannot surpass the performance of the BERT model. Thus, we employ this fine-tuned BERT model for categorizing all the tweets from the remaining dataset. Having employed BERT on the remaining dataset, we get a refined dataset of the four categories of tweets spreaded across the three stages as shown in Table~\ref{tab2}.

We also calculate the confusion matrix for our best performing model BERT as shown in Fig~\ref{fig3}. As can be seen from the confusion matrix, we obtain an excellent classification performance ($>$0.90) for stigmatization and non-racism classification categories, a higher performance ($>$0.85 while $<$0.90) for the exclusion category, and moderately higher performance ($>$0.75 while $<$0.85) for the other two remaining categories of offensiveness and blame.

\subsection{Topic modelling}
Topic modelling is one of the most extensively used methods in natural language processing for finding relationships across text documents, topic discovery and clustering, and extracting semantic meaning from a corpus of unstructured data \citep{jelodar2019latent}. Many techniques have been developed by researchers such as Latent Semantic Analysis (LSA) \citep{deerwester1990indexing}, Probabilistic Latent Semantic Analysis (pLSA) \citep{hofmann1999probabilistic} for extracting semantic topic clusters from the corpus of data. In the last decade, Latent Dirichlet Allocation (LDA) \citep{blei2003latent} has become a successful and standard technique for inferring topic clusters from texts for various applications such as opinion mining \citep{zhai2011constrained}, social medial analysis \citep{cohen2013classifying}, event detection \citep{lin2010pet} and consequently there have also been various developed variants of LDA \citep{blei2010supervised} and \citep{blei2003hierarchical}.

For our research, we adopt the baseline LDA model with Variational Bayes sampling from Gensim\footnote{https://pypi.org/project/gensim/} and the LDA Mallet model \citep{mccallum2002mallet} with Gibbs sampling for extracting the topic clusters from the text data. Before passing the corpus of data to the LDA models, we perform data pre-processing and cleaning which include the following steps. Firstly, we remove any new line characters, punctuations, URLs, mentions and hashtags. Later we tokenize the texts in the corpus and also remove any stopwords using the Gensim utility of pre-processing and stopwords defined in the NLTK\footnote{https://pypi.org/project/nltk/} corpus. Finally, we make bigrams and lemmatize the words in the text.

After employing the above pre-processing for our corpus, we employ topic modelling using LDA from Gensim and LDA Mallet. We perform experiments by varying the number of topics from 5 to 25 at an interval of 5 and checking the corresponding coherence score of the model \citep{fang2016exploring}.  We train the models for 1000 iterations with varying number of topics, optimizing the hyperparameters every 10 passes after each 100 pass period. We set the values of $\alpha$, $\beta$ which control the distribution of topics and the vocabulary words amongst the topics to the default settings of 1 divided by the number of topics. We notice from our experiments that LDA Mallet has a higher coherence score (0.60-0.65) compared to the LDA model from Gensim (0.49-0.55) and thus we select LDA Mallet model for the task of topic modelling on our corpus of data.

The above strategy is employed for each racist and xenophobic category and for every stage individually. We find the highest coherence score corresponding to a specific number of topics for each category and stage. To analyse the results, we reduce the number of topics to 5 by clustering closely related topics using equation~\ref{eq:clustering}.

where $N$ refers to the number of topics to be clustered, $M$ represents the number of keywords in each topic, $p_j$ corresponds to the probability of the word $x_i$ in the topic, and $T_c$ is the resultant topic containing the average probabilities of all the words from the $N$ topics. We then represent the top 10 highest probability words in the resultant topic for every category and stage as is shown in Tables~\ref{tab4} to~\ref{tab7}.

\begin{equation}
T_{c}=\dfrac{\left( \sum ^{N}_{i=1}\sum ^{M}_{j=1}p_{j}x_{j}\right) }{N}
\label{eq:clustering}
\end{equation}

\section{Findings}\label{sec11}

Table~\ref{tab3} illustrates the distribution of racist tweets of the four categories across the three stages. Tables~\ref{tab4},~\ref{tab5},~\ref{tab6}, and~\ref{tab7} demonstrate the ten most salient terms related to the generated five topics for each stage (S1, S2, and S3) of four categories. Each topic was summarized through the correlation between the ten terms. We put a question mark for topics from which no pattern can be generated. The below section provides a detailed analysis of the dynamics of the four categories.

\begin{table}[h]
\centering
\caption{Distribution of tweets amongst the four categories across the three stages.}\label{tab3}%
\resizebox{0.6\textwidth}{!}{\begin{tabular}{lcccc}
\hline
Category & Total & S1 & S2 & S3 \\
\hline
Stigmatization & 116584 & 3723 & 5687 & 107174 \\
Offensiveness & 10503 & 1722 & 1808 & 6973\\
Blame & 39765 & 31 & 777 & 38957\\
Exclusion & 10293 & 872 & 1341 & 8080 \\
\hline
\end{tabular}}
\end{table}

\subsection{Stigmatization}\label{subsec111}

\begin{table*}[!t]
\centering
\caption{Extracted topics and their corresponding keywords for the category of stigmatization spread across the three stages S1, S2, and S3.}
\resizebox{\textwidth}{!}{
\begin{tabular}{p{5mm}|l|cccccccccc}
\hline
\multirow{5}{*}{S1} & T1.\textbf{Virus} & virus & spread & country & travel & year & control & \textit{chinese} & ban & corona & show\\

& T2.\textbf{China/Chinese} & \textit{chinese} & virus & deadly & \textit{china} & situation & mask & stop & animal & source & eat\\

& T3.\textbf{Infection} & people & case & health & infect & confirm & death & sar & number & report & market\\

& T4.\textbf{Outbreak} & \textit{china} & coronavirus & wuhan & outbreak & city & hospital & news & patient & put & state\\

& T5.\textbf{Travel} & world & \textit{china} & government & make & people & time & day & bad & flight & start\\
\hline
\multirow{5}{*}{S2} & T1.\textbf{Emergency} & virus & spread & day & year & corona & show & emergency & food & kit & supply\\

& T2.\textbf{Globe} & \textit{china} & world & time & country & report & death & global & health & travel & confirm\\

& T3.\textbf{Infection} & people & case & call & ncov & infect & kill & pack & state & flu & number\\

& T4.\textbf{China} & \textit{china} & coronavirus & wuhan & outbreak & quarantine & stop & find & man & dead & thing\\

& T5.\textbf{Chinese} & \textit{chinese} & make & mask & government & news & good & work & citizen & start & respirator\\
\hline
\multirow{5}{*}{S3} & T1.\textbf{Government} & \textit{china} & world & spread & country & lie & pay & communist & government & ccp & make\\

& T2.\textbf{?} & time & make & \textit{india} & good & give & work & day & back & fight & buy\\

& T3.\textbf{China} & \textit{china} & coronavirus & case & death & covid & country & economy & war & number & wuhan\\

& T4.\textbf{Chinese} & \textit{chinese} & virus & people & call & stop & racist & start & die & blame & corona\\

& T5.\textbf{US} & \textit{american} & trump & \textit{state} & medium & president & \textit{america} & news & great & propaganda & show\\
\hline

\end{tabular}
}
\label{tab4}
\end{table*}

According to Table~\ref{tab3}, stigmatization continuously acted as the dominant racist theme across the three early stages of COVID-19 (S1:3723; S2:5687; S3:107174). Based on the result generated from topic modelling (see Table~\ref{tab4}), the main topics related to stigmatization at the first stage included “virus”, “China/Chinese”, “infection”, “outbreak” and “travel”. In general, stigmatization of this stage focused on the infectious nature of the virus, the association between the virus and China/Chinese, and the outbreak of the virus. Especially, the sub-topics under “China/Chinese” included “mask”, “animal”, and “eat”, which echoed the heated debates around China and Chinese at that time – to specify, whether wearing a mask that was advocated by China government would be helpful; and whether the origin of this virus was associated with the eating habits of Chinese people. 

At the second stage, the leading topics changed to “emergency”, “globe”, “infection”, “China”, and “Chinese”. First, we noticed that due to the global outbreak of COVID-19 at this stage, expression of stigmatization started to pay more attention to the world situation. In addition, “China” and “Chinese” were separated and became two main topics of stigmatization. To specify, the stigmatization around “China” included sub-topics such as “wuhan”, “quanrantine”, “dead”, which reflected the attention drawn to the status of Wuhan. The stigmatization around “Chinese” was more likely to focus on the government, with sub-topics such as “mask”, “government”, “citizen”, and “news”. 

Main stigmatization topics at the third stage included “government”, “China”, “Chinese”, “US”, and one topic that could not be identified due to its irrelevant sub-topics. Notably, “government” has become an independent topic of stigmatization at this stage. Especially, under the theme of “government”, “communist” and “ccp” were co-existed with “lie”.  In addition, “US” emerged as a new topic, and Trump who was the president of the United States during that time became one critical sub-topic under “US”.  This might happen after Trump’s twitter that referred to COVID-19 as Chinse virus. Also, it is worth noting that the conflation between virus and a race/ethnicity contributed to a rapid growth of stigmatization-oriented racist opinions from stage 2 to stage 3 (from 5687 to 107174).

\subsection{Offensiveness}\label{subsec112}

\begin{table*}[!t]
\centering
\caption{Extracted topics and their corresponding keywords for the category of offensiveness spread across the three stages S1, S2, and S3.}
\resizebox{\textwidth}{!}{\begin{tabular}{p{5mm}|l|cccccccccc}
\hline
\multirow{5}{*}{S1} & T1.\textbf{?} & country & ccp & citizen & virus & arrest & live & system & security & foreign & understand\\

& T2.\textbf{Government} & people & government & democracy & support & life & year & regime & \textit{uyghur} & camp & give\\

& T3.\textbf{?} & \textit{china} & world & spread & stop & communist & happen & \textit{taiwan} & wuhan & govt & ban\\

& T4.\textbf{Muslim} & \textit{chinese} & make & muslim & good & kill & police & terrorist & bad & party & lie\\

& T5.\textbf{Human right} & world & freedom & \textit{hong\_kong} & human & human\_right & time & free & stand & \textit{hk} & fight\\
\hline
\multirow{5}{*}{S2} & T1.\textbf{Freedom} & world & stop & freedom & truth & spread & good & free & \textit{hk} & speech & life\\

& T2.\textbf{Ccp} & \textit{china} & \textit{chinese} & ccp & virus & happen & \textit{wuhan} & evil & communist & time & \textit{uyghur}\\

& T3.\textbf{People} & people & make & kill & lie & ppl & trust & camp & police & thing & man\\

& T4.\textbf{China} & \textit{china} & country & regime & pay & money & outbreak & start & work & force & control\\

& T5.\textbf{Human right} & government & citizen & human & fight & support & hong\_kong & taiwan & give & democracy & death\\
\hline
\multirow{5}{*}{S3} & T1.\textbf{Death} & world & people & pay & lie & kill & truth & fight & life & die & humanity\\

& T2.\textbf{Government} & time & call & government & \textit{india} & communist & pandemic & give & global & send & real\\

& T3.\textbf{Virus} & \textit{chinese} & virus & spread & \textit{wuhan} & corona & product & buy & control & big & day\\

& T4.\textbf{China} & \textit{china} & country & make & ccp & stop & good & coronavirus & human & trust & support\\

& T5.\textbf{World} & \textit{china} & world & war & case & start & covid & economy & death & state & \textit{italy}\\
\hline
\end{tabular}}
\label{tab5}
\end{table*}

At the first and second stages, offensiveness was the second-most-frequently-mentioned theme of racist tweets (S1: 1722; S2: 1808, see Table~\ref{tab3}). To specify, according to Table~\ref{tab5}, at the first stage, the main topics of offensiveness included “government”, “muslim”, and “human right”, in addition to two unidentified themes. Under “government”, we discovered that some sub-topics were not directly related to COVID-19. Instead, they (e.g., “uyghur” and “camp”) were more likely to target the sensitive internal affairs of China. Similarly, the topics of “muslim” and “human right” also tended to emphasize the China’s internal affairs that had been heatedly discussed before COVID-19. While the offensive sub-topics under Muslim included “kill”, “police”, “terrorist”, “bad”, “party”, and “lie”, the discussion on “human right” centred on “freedom” and “hongkong”.

Offensive language at the second stage still targeted China's internal affairs and political system. The main topics included “freedom”, “ccp”, “people”, “China”, and “human right”. First, “freedom” emerged as a new topic at this stage. Second, the political attack became more specified, transferring from the “government” to “ccp”. And the “ccp” related discussion still focused on “uyghur”. But “wuhan” became a new sub-topic under “ccp”. “Human right” was still a major topic. However, under “human right”, besides “hongkong”, “taiwan” became a new sub-topic.

At the third stage, the main topics of offensiveness changed to “death”, “government”, “virus”, “China”, and “world”. Under all topics, “uyghur”, “hongkong”, and “taiwan” were out of the picture. This indicated a shifting focus of offensiveness that started to shed more illumination on the virus rather than the political debates around China’s internal affairs. This theme shift was accompanied by the reduced attention to offensive expression. According to Table~\ref{tab3} (S3: 6973), offensiveness became the least important theme of racist tweets. In general, we find that when COVID-19 was reported to be discovered in China, offensiveness was largely deployed to raise hatred by relating this virus with China’s internal affairs. However, alongside the global outbreak of COVID-19, less attention was drawn to these internal affairs. In the meanwhile, fewer opinions were expressed in an offensive way. 

\subsection{Blame}\label{subsec113}

\begin{table*}[!t]
\centering
\caption{Extracted topics and their corresponding keywords for the category of blame spread across the three stages S1, S2, and S3.}
\resizebox{\textwidth}{!}{\begin{tabular}{p{5mm}|l|cccccccccc}
\hline
\multirow{5}{*}{S1} & T1.\textbf{Lie} & lie & spread & virus & autocracy & deceit & imagine & true & horrible & infect & country\\

& T2.\textbf{Death} & \textit{china} & dead & die & day & order & monstrosity & true & thing & kong & high\\

& T3.\textbf{Safety} & coronavirus & move & lot & cvirus & epicenter & safety & march & careful & knowingly & health\\

& T4.\textbf{Time} & \textit{wuhan} & lunar\_new & sick & year & time & absolutely & medium & mutate & emperor & truth\\

& T5.\textbf{Infection} & people & \textit{chinese} & make & online & pandemic & catch & number & infect & community & official\\
\hline
\multirow{5}{*}{S2} & T1.\textbf{Government} & lie & \textit{chinese} & coronavirus & government & \textit{wuhan} & cover & day & body & thing & care\\

& T2.\textbf{Spread} & world & country & spread & happen & trust & kill & threat & steal & dead & face\\

& T3.\textbf{China} & \textit{china} & truth & bad & free & money & communist & case & find & start & move\\

& T4.\textbf{Virus} & virus & stop & make & control & good & \textit{china} & fight & live & report & human\\

& T5.\textbf{Death} & people & time & number & die & real & life & entire & back & citizen & death\\
\hline
\multirow{5}{*}{S3} & T1.\textbf{World} & world & \textit{china} & country & pay & pandemic & kill & global & economy & war\\

& T2.\textbf{?} & people & stop & human & \textit{american} & eat & put & president & market & happen & live\\

& T3.\textbf{Lie} & \textit{china} & lie & coronavirus & \textit{wuhan} & blame & die & case & cover & truth & number\\

& T4.\textbf{?} & make & time & \textit{china} & good & start & buy & trust & back & thing & country\\

& T5.\textbf{Government} & \textit{chinese} & virus & \textit{china} & government & call & communist & ccp & covid & spread & hold\\
\hline

\end{tabular}}
\label{tab6}
\end{table*}

According to Table~\ref{tab3}, tweets for blaming grew rapidly across the stages of COVID-19 (S1:31; S2 777; S3: 38957). Especially, at the first two stages, blame only occupied the smallest number of racist tweets. However, at the third stage, blame became the second leading theme following stigmatization.
To specify, according to Table~\ref{tab6}, at the first stage, the main topics to “blame” included “lie”, “death”, “safety”, “time”, and “infection”. We found that blaming reactions at this stage tended to target the negative consequence of COVID-19. For instance, as noted, “stigmatization” tended to associate lie with the political system. However, “blame” was more likely to focus on the consequence of “lie” such as “spread”, “deceit”, “horrible”, and “infect”. This focus can also be easily detected from the rest four topics - “death”, “safety”, “time”, and “infection” that are variously related to the threats brought about by COVID-19 to people’s health and safety. 

The second stage of “blame” involved “government”, “spread”, “china”, “virus”, and “death” as the main topics. It is important to note that, besides the words describing the negative COVID-19 consequence, “government” and “china” emerged as two new topics. These two new topics indicated that racist reactions tended to increasingly blame COVID-19 on China and its management of COVID-19. Especially, “government” was associated with “lie”, and “china” was associated “truth”. This suggested that there might be an increasing number of tweets blaming China and government for telling lies and regarding the lying behaviour as the major reason resulting in the outbreak of COVID-19. 
	
Main topics at the third stage included “world”, “lie”, “government”, and two unidentified topics. Akin to the second stage, “lie” and “government” indicated that racist tweets still tended to lay the blame on the “lie” of “government”. Notably, at this stage, “world” emerged as a new main topic. In the category of “world”, we found sub-topics such as “kill”, “global”, “economy”, “war”, “china”, and “pay”, which indicated that “blame” might have been leveraged to emphasize the negative effects brought about by COVID-19 to the world. 

In general, across the three stages of COVID-19, the escalation of blaming behaviours was accompanied by an increasingly specified target to blame. In addition, this target was blamed to contribute to COVID-19 as well as its constantly expanded negative influence across the globe. In so doing, as revealed in prior research, the blaming reactions continued to reinforce the processes of “othering” \citep{bailey2005racialised} to draw the boundary between different racial groups.

\subsection{Exclusion}\label{subsec114}

\begin{table*}[!t]
\centering
\caption{Extracted topics and their corresponding keywords for the category of exclusion spread across the three stages S1, S2, and S3.}
\resizebox{\textwidth}{!}{\begin{tabular}{p{5mm}|l|cccccccccc}
\hline
\multirow{5}{*}{S1} & T1.\textbf{Government} & support & gov & join & people & evil & time & stand & sanction & government & money\\

& T2.\textbf{Human right} & product & world & stop & human\_right & freedom & tag & good & challenge & ppl & economic\_infiltration\\

& T3.\textbf{Boycott} & \textit{china} & \textit{hong\_kong} & fight & regime & boycott & show & international & control & trust & communist\\

& T4.\textbf{Trade} & make & buy & ccp & day & thing & friend & \textit{taiwan} & \textit{japan} & hope & today\\

& T5.\textbf{Virus} & country & \textit{chinese} & people & spread & year & human & animal & protect & virus & eat\\
\hline
\multirow{5}{*}{S2} & T1.\textbf{Nation} & people & \textit{chinese} & animal & happen & government & initiative & nation & show & economy & law\\

& T2.\textbf{Virus} & virus & control & truth & support & live & kill & boycott & start & stand & cover\\

& T3.\textbf{Threat} & \textit{china} & time & lie & threat & company & trust & big & entire & spy & \textit{wuhan}\\

& T4.\textbf{Human right} & world & country & freedom & spread & human\_right & economic & thing & evil & steal & raise\\

& T5.\textbf{Trade} & make & product & stop & buy & day & \textit{china} & good & ccp & challenge & coronavirus\\
\hline
\multirow{5}{*}{S3} & T1.\textbf{Virus} & \textit{china} & virus & world & pay & spread & ccp & covid & corona & market & call\\

& T2.\textbf{Pandemic} & world & \textit{china} & company & communist & coronavirus & pandemic & global & nation & trust & war\\

& T3.\textbf{Trade} & \textit{chinese} & make & product & buy & boycott & stop & good & \textit{India} & economy & \textit{Indian}\\

& T4.\textbf{Human right} & people & lie & government & human & life & back & animal & kill & eat & bring\\

& T5.\textbf{China} & \textit{china} & country & time & start & business & give & thing & app & sell & money\\
\hline

\end{tabular}}
\label{tab7}
\end{table*}

According to Table~\ref{tab3}, exclusion remained as the second least mentioned theme of racist expression on twitter across the three stages (S1: 872; S2: 1341: S3:8080). However, the topics of exclusion kept on changing. To specify, according to Table~\ref{tab7}, at the first stage, the main topics include “government”, “human right”, “boycott”, “trade”, and “virus”.  Like other themes, exclusion also targeted political system of China government and its management of COVID-19. However, exclusion also included topics such as “boycott” and “trade”. “Boycott” indicated the purpose of exclusion that was expected to lead to a rejection, while “trade” specified exclusion in an economic way.

At the second stage, exclusion topics became “nation”, “virus”, “threat”, “human right”, and “trade”. Amongst the five topics, “nation” and “threat” were new. It is interesting to note that an extended scope of exclusion that has been transferred from “government” to “nation.” Second, “threat”, as a new focus of exclusion, included sub-topics such as “lie”, “trust”, “spy”, which suggested the unexpected threat that may arise from distrust had become an important reason for exclusion.

At the third stage, topics changed to “virus”, “world”, “trade”, “human right”, and “China.” Two new topics included “world” and “China”. “World” suggested an increasingly globalized discussion on exclusion, including sub-topics such as “global”, “nation”, “trust”, and “war”. Opinion expression concerning “China” tended to place the emphasis on the discussion on the business of China, including sub-topics such as “country”, “business”, “app”, “sell”, and “money”. 

Different from other categories, exclusion paid more attention to trade and economy. This suggested that the expression of exclusion tended to focus on the economic aspects between China and the world. Additionally, boycott emerged as a new theme especially in the early stage suggesting that early approach of the discussion might focus on abandoning Chinese manufactured products and putting less reliance on China. Moreover, the reason behind exclusion seems to be the feeling of threat that tends to be originated from the distrust on China government. 

\section{Discussion and conclusions}\label{sec12}
Our study makes unique contribution that enriches the scholarly discussion on deviant racist behaviours on social media. First, bridging computational methods with social science theories, we transcend a binary classification of racist tweets and instead, propose a multi-dimensional model for racism detection, classification, and analysis. This method, echoing the complicated and dynamic nature of racism, maps the evolvement of racist behaviours alongside the development of COVID-19. Furthermore, the multi-dimensional categorization of racist behaviours also enables the capturing of the diversity of topics – how different focuses of racist tweets fell under different categories, and how the discussion focus of each category kept on changing across time. 

This leads to the second contribution that lies in policy implementation. To specify, the nuanced and dynamic understanding of the racist reactions in the context of COVID-19 will enable the policy makers to have a better interpretation of the possible motivations driving the racist reactions. For instance, as our findings revealed, compared to offensiveness, blame, and exclusion, stigmatization was more likely to act as the leading factor triggering the racist behaviours. Another example is that offensive language was normally deployed to attack the internal affairs of China which might be irrelevant to COVID-19. Better knowledge regarding the reasons behind the public racist reactions could lead to the enactment of more effective policies to prevent the escalation of the race-related deviant behaviours and hate speech.

Additionally, the stage-wise analysis contributes to the enactment of intervention policies with a more specified target at different stages of pandemic. For instance, at the third stage, blame became the most rapidly growing theme, while less and less people were interested in using offensive language. Therefore, the intervention policy can change the focus accordingly across the stages for a better detection and monitoring of racist posts on social media platforms. 
	
Lastly but not least, our study contributes to providing insights into the possible route for interdisciplinary research in the domain of social network analysis and mining. Especially, our study points a direction of deploying social science theories to develop the computational methods for big social data analytics. Future research can also consider embracing the social science perspectives to advance the detection and analysis of linguistic features concerning a particular topic.


\bibliography{sn-bibliography.bib}


\end{document}